\documentclass[aps,prd,amsmath
,nofootinbib         
,twocolumn         
]{revtex4}
\usepackage{amsmath}
\usepackage{amsfonts}
\usepackage{amssymb}

\allowdisplaybreaks[4]     

\begin{document}

\title{New classes of exact interior nonvacuum solutions to the GR field equations for spacetimes sourced by a rigidly rotating stationary cylindrical anisotropic fluid}

\author{Marie-No\"{e}lle C\'{e}l\'{e}rier}
\affiliation{Laboratoire Univers et TH\'eories, LUTH, Observatoire de Paris, Universit\'e PSL, CNRS, Universit\'e de Paris, 5, Place Jules Janssen, 92190 Meudon, France}
\email{marie-noelle.celerier@obspm.fr}

\date{4 September 2021}

\begin{abstract}

A double new class of solutions to the general relativity field equations describing interior spacetimes sourced by stationary cylindrical anisotropic fluids with principal stress directed along the symmetry axis is displayed. These solutions are required to satisfy regularity and junction conditions so that they can be possibly used to represent rotating astrophysical objects. Mathematical and physical properties are analyzed. The spacetime two independent parameters are physically interpreted, and they are shown to define two different solution classes together with stating the latters' properties.

\end{abstract}

\maketitle

\section{Introduction} \label{intro}

Cylindrical symmetry implying two Killing vectors has attracted much attention since the pioneering 1919 work by Levi-Civita identifying static vacuum cylindrical spacetimes \cite{LC19}. Their extension to the stationary case exhibiting three Killing vectors, was obtained in 1924 by Lanczos \cite{L24} who considered a rigidly rotating infinite dust cylinder with and without a cosmological constant, and by Lewis in 1932 for vacuum \cite{L32}. The Lanczos solution was independently rediscovered by van Stockum in 1937 for the zero cosmological constant case \cite{vS37}, and this solution is often improperly attributed to this single author. The Lewis solution describes a vacuum exterior gravitationally sourced by a matter cylinder rotating around its symmetry axis. Depending on whether the constant parameters appearing in the metric functions are real or complex, the solutions are said to belong to the Weyl class or to the Lewis class, respectively. The vacuum solution outside a cylindrical source in translation along its symmetry axis is mathematically akin to the Lewis solution with exchanged $z$ and $\phi$ coordinates. They are however physically different \cite{C19}. Now, cylindrically symmetric spacetimes have been extensively investigated for a number of different purposes \cite{G09,S09}. For a recent review on cylindrical systems in general relativity (GR), see \cite{B20}.

In \cite{D06}, the rigid rotation of nonvacuum stationary spacetimes sourced by a cylindrical anisotropic fluid has been considered. In \cite{CS20}, this study has been extended to the nonrigid rotation case, while the rigid case analysis proposed in \cite{D06} has been supplemented, with a focuss on its Weyl tensor gravitoelectromagnetic properties. Now, as an existence proof for solutions to constraint equations appearing in this study, an example of particular rigidly rotating metric has been incidentally proposed there. However, it is easy to check that this solution is trivial. A further analysis yields a constraint on the $h(r)$ basic building function forcing it to be constant, and this solution to be therefore both flat and static, i.e., Minkowski: see the Appendix in \cite{CS20}. Note, however, that the other results displayed in \cite{CS20} are not precluded by this observation. In the present work, an improved solution to the field equations for the same fluid equation of state is exhibited and studied. This new solution is exact and is thoroughly examined here while a number of its mathematical and physical  features are analyzed. 

The equation of state studied here is a generalization of that considered in \cite{D06} and the same as one of those analyzed in \cite{CS20}. To the author's knowledge, these works together with the one displayed in the present paper constitute the first attempt to introduce anisotropy into rigidly rotating stationary cylindrical spacetimes. Even though the equation of state proposed here is rather simple, as it is generally the case when one starts a new path exploration, it might anyhow prove possibly useful as an approximation to improve our cosmological and astrophysical bestiary knowledge. A couple of such proposals are displayed in this paper. In any case, this exact solution to a new physical setup will certainly allow us to increase our understanding of cylindrically symmetric fluids in GR.

The paper is organized as follows: In Sec. \ref{ci}, the stationary cylindrically symmetric line element which will be used for the present purpose is set up. In Sec. \ref{rrme}, the new exact solution is constructed from the field equations pertaining to the problem, through its regularity and junction conditions, up to its final form. Important mathematical and physical properties pertaining to this solution are analyzed in Sec. \ref{physprop} where, in particular, two different classes of solutions are disclosed. The conclusions are displayed in Sec. \ref{concl}.

\section{Cylindrical spacetime inside the source} \label{ci}

Consider a rigidly rotating stationary cylindrically symmetric anisotropic nondissipative fluid bounded by a cylindrical surface $\Sigma$ whose principal stresses $P_r$, $P_z$ and $P_\phi$ obey the equation of state $P_r= P_{\phi}=0$. Its stress-energy tensor can thus be written as
\begin{equation}
T_{\alpha \beta} = \rho V_\alpha V_\beta + P_z S_\alpha S_\beta, \label{setens}
\end{equation}
where $\rho$ is the fluid energy density, $V_\alpha$ is the timelike fluid 4-velocity, and $S_\alpha$ is a spacelike 4-vector satisfying
\begin{equation}
V^\alpha V_\alpha = -1, \quad  S^\alpha S_\alpha = 1, \quad V^\alpha S_\alpha = 0. \label{fourvec}
\end{equation}
We assume, for the inside $\Sigma$ spacetime, the spacelike $\partial_z$ Killing vector to be hypersurface orthogonal, such as to ease its subsequent matching to the exterior Lewis Weyl class metric. Hence, the stationary cylindrically symmetric line element reads
\begin{equation}
\textrm{d}s^2=-f \textrm{d}t^2 + 2 k \textrm{d}t \textrm{d}\phi +\textrm{e}^\mu (\textrm{d}r^2 +\textrm{d}z^2) + l \textrm{d}\phi^2, \label{metric}
\end{equation}
where $f$, $k$, $\mu$, and $l$ are real functions of the radial coordinate $r$ only. Owing to cylindrical symmetry, the coordinates conform to the following ranges:
\begin{eqnarray}
-\infty &\leq& t \leq +\infty, \quad 0 \leq r, \quad -\infty \leq z \leq +\infty, \nonumber \\ 
0 &\leq& \phi \leq 2 \pi, \label{ranges}
\end{eqnarray}
where the two limits of the $\phi$ coordinate are topologically identified. These coordinates are numbered $x^0=t$, $x^1=r$, $x^2=z$, and $x^3=\phi$.

\section{The rigidly rotating new solution} \label{rrme}

For rigid rotation where a corotating frame is chosen for the stationary fluid source \cite{D06,CS20}, the fluid 4-velocity can be written as
\begin{equation}
V^\alpha = v \delta^\alpha_0, \label{r4velocity}
\end{equation}
where $v$ is a function of $r$ only. The timelike condition for $V^\alpha$ provided in (\ref{fourvec}) thus reads
\begin{equation}
fv^2 = 1. \label{timelike}
\end{equation}
The spacelike 4-vector used to define the stress-energy tensor and verifying conditions (\ref{fourvec}) can be chosen as
\begin{equation}
S^\alpha = \textrm{e}^{-\mu/2}\delta^\alpha_2, \label{salpha}
\end{equation}
and a calculation intermediate function $D$ is defined as
\begin{equation}
D^2 = fl + k^2. \label{D2}
\end{equation}

\subsection{Field equations} \label{fe}

With the above choice for the two 4-vectors defining the stress-energy tensor, and using (\ref{r4velocity})--(\ref{D2}) into (\ref{setens}), one obtains the stress-energy tensor components corresponding to the five nonvanishing Einstein tensor components, and one can thus write the following five field equations for the inside $\Sigma$ spacetime. 

With $h(r)$ defined as $h(r)\equiv P_z(r)/\rho(r)$, they read
\begin{eqnarray}
G_{00} &=& \frac{\textrm{e}^{-\mu}}{2} \left[-f\mu'' - 2f\frac{D''}{D} + f'' - f'\frac{D'}{D} + \frac{3f(f'l' + k'^2)}{2D^2}\right] \nonumber \\
&=& \kappa\rho f, \label{G00}
\end{eqnarray}
\begin{eqnarray}
G_{03} &=&  \frac{\textrm{e}^{-\mu}}{2} \left[k\mu'' + 2 k \frac{D''}{D} -k'' + k'\frac{D'}{D} - \frac{3k(f'l' + k'^2)}{2D^2}\right]  \nonumber \\
&=& - \kappa\rho k, \label{G03}
\end{eqnarray}
\begin{eqnarray} 
G_{11} &=& \frac{\mu' D'}{2D} + \frac{f'l' + k'^2}{4D^2} = 0, \label{G11}
\end{eqnarray}
\begin{eqnarray}
G_{22} &=& \frac{D''}{D} -\frac{\mu' D'}{2D} - \frac{f'l' + k'^2}{4D^2} = \kappa \rho h \textrm{e}^\mu, \label{G22}
\end{eqnarray}
\begin{eqnarray}
G_{33} &=&  \frac{\textrm{e}^{-\mu}}{2} \left[l\mu'' + 2l\frac{D''}{D} - l'' + l'\frac{D'}{D} - \frac{3l(f'l' + k'^2)}{2D^2}\right] \nonumber \\
&=& \kappa \rho \frac{k^2}{f}, \label{G33}
\end{eqnarray}
where the primes stand for differentiation with respect to $r$.

\subsection{Stress-energy tensor conservation }

Writing the stress-energy tensor conservation is analogous to writing the Bianchi identity
\begin{equation}
T^\beta_{1;\beta} = 0. \label{Bianchi}
\end{equation}
From (\ref{setens}), we have
\begin{equation}
T^{\alpha \beta} = \rho V^\alpha V^\beta + P_z S^\alpha S^\beta, \label{setenscontra}
\end{equation}
with $V^\alpha$ given by (\ref{r4velocity}), and the spacelike vector $S^\alpha$ given by (\ref{salpha}), which can be inserted into (\ref{setenscontra}). Then, using (\ref{metric}) and (\ref{timelike}), the Bianchi identity (\ref{Bianchi}) reduces to
\begin{equation}
T^\beta_{1;\beta} = \frac{1}{2} \rho \frac{f'}{f} - \frac{1}{2} P_z \mu'  = 0, \label{Bianchi2}
\end{equation}
or, inserting the $h(r)$ function defined above,
\begin{equation}
 \frac{f'}{f} - h \mu'  = 0. \label{Bianchi3}
\end{equation}

\subsection{Solving the field equations} \label{solving}

In \cite{CS20}, a partly integrated equation for a rigidly rotating $P_r= P_{\phi}=0$ fluid has been derived as Eq.(152) proceeding solely from the field equations and the Bianchi identity. Two solutions of this Eq.(152) have been identified there. The second solution, Eq.(175) in \cite{CS20}, is considered here. 

Since only five independent differential equations are available for six unknowns, i.e., the four metric functions $f$, $k$, $\textrm{e}^\mu$, and $l$, the energy density $\rho$, and the pressure defined either by $P_z$ or by $h$ -- this last option will be retained here -- the equation set needs to be closed by an additional one. The particular assumption displayed in \cite{CS20} as Eq. (111) is chosen for this purpose:
\begin{equation}
\mu' = \frac{2h'}{1-h} + \frac{2h'}{h}, \label{sol1}
\end{equation}
which can be integrated as
\begin{equation}
\textrm{e}^\mu = c_{\mu} \frac{h^2}{(1-h)^2}, \label{sol2}
\end{equation}
where $c_{\mu}$ is an integration constant.
Inserting (\ref{sol1}) into the Bianchi identity (\ref{Bianchi3}), one obtains
\begin{equation}
\frac{f'}{f} = \frac{2h'}{1-h}, \label{sol3}
\end{equation}
which can be integrated as
\begin{equation}
f =  \frac{c_f}{(1-h)^2}, \label{sol4}
\end{equation}
$c_f$ being another integration constant.

Now, (\ref{G00}) combined with (\ref{G03}) can be written as
\begin{equation}
\left(\frac{kf' - fk'}{D}\right)' = 0, \label{sol5}
\end{equation}
which can be integrated as \cite{D06}
\begin{equation}
kf' - fk' = 2c D, \label{sol6}
\end{equation}
where $2c$ is an integration constant, the factor 2 being chosen here for further convenience. Considered as a first-order ordinary differential equation for $k(r)$, (\ref{sol6}) possesses as a general solution
\begin{equation}
k = f \left(c_0 - 2c\int_{r_0}^r \frac{D(v)}{f(v)^2} \textrm{d}v \right), \label{sol7}
\end{equation}
where $c_0$ and $r_0$ are new integration constants, and which, with expression (\ref{sol4}) for $f$ inserted, can be written as
\begin{equation}
k = \frac{c_f}{(1-h)^2} \left[c_0 - \frac{2c}{c_f^2}\int_{r_0}^r \left(1-h(v)\right)^4 D(v) \textrm{d}v \right]. \label{sol8}
\end{equation}
The last metric function $l$ thus follows from (\ref{D2}) as
\begin{eqnarray}
l &=& \frac{(1-h)^2}{c_f}\left\{D^2 - \frac{c_f^2}{(1-h)^4} \right. \nonumber \\
&\times& \left.  \left[c_0 - \frac{2c}{c_f^2}\int_{r_0}^r \left(1-h(v)\right)^4 D(v) \textrm{d}v \right]^2 \right\}. \label{sol9}
\end{eqnarray}

The field equations (\ref{G00}), (\ref{G11}), and (\ref{G22}), as well as, equivalently, (\ref{G03})--(\ref{G22}) give an expression for $D'/D$ as a function of $h$ and of its first and second derivatives which reads
\begin{eqnarray}
\frac{D'}{D} &=& \frac{1}{3+h} \left[ - (1-h)\frac{h''}{h'} + (1-2h)\frac{h'}{h} \right. \nonumber \\
&+& \left. \frac{(1+h)h'}{1-h} -\kappa c_{\mu}\rho \frac{(1+h)h^3}{(1-h)h'} \right].  \label{sol10}
\end{eqnarray} 

Now, inserting (\ref{sol2}), (\ref{sol4}),  (\ref{sol8})--(\ref{sol10}) into the field equation (\ref{G11})  gives
\begin{eqnarray}
\rho &=& \frac{1-h}{\kappa c_{\mu} h^2(1+h)} \left[ - \frac{(1-h)}{h}h'' + \frac{(1-2h^2)}{h^2(1+h)}h'^2 \right. \nonumber \\
&+& \left. \frac{c^2}{c_f^2}\frac{(1-h)^5(3+h)}{1+h} \right].  \label{sol11}
\end{eqnarray}
Then inserting the same expressions into (\ref{G00}), or equivalently into (\ref{G33}), one obtains
\begin{eqnarray}
\rho &=& \frac{2}{3 \kappa c_{\mu}} \left[ \frac{2(1-h)}{h^2(1+h)}h'' + \frac{(1+h+4h^2)}{h^3(1+h^2)}h'^2 \right. \nonumber \\
&+& \left. \frac{2 c^2}{c_f^2}\frac{(1-h)^6(3+h)}{h^2(1+h)^2} \right].  \label{sol12}
\end{eqnarray}
Equalizing both expressions for $\rho$ given by  (\ref{sol11}) and  (\ref{sol12}) yields a second-order ordinary differential equation for $h$ that reads
\begin{equation}
h'' = \frac{(1-2h-2h^2)}{h(1-h)(1+h)}h'^2 - \frac{c^2}{c_f^2}\frac{h(1-h)^5}{1+h}, \label{sol13}
\end{equation}
whose general solution is
\begin{equation}
c_r + r = \frac{c_{\gamma}}{ c_{\alpha}^{\frac{1}{2}}}\int_{h_0}^{h(r)} \left[\frac{1+u}{u^2(1-u)^3(- 2\ln u + 4u - u^2 + c_{\beta})}\right]^\frac{1}{2} \textrm{d}u, \label{sol14}
\end{equation}
where $c_r$, $c_{\alpha}$, $c_{\beta}$, and $c_{\gamma}$ are integration constants and where a $\pm$ sign appearing in the calculations has been absorbed into the $c_{\gamma}$ definition. From (\ref{sol14}), the $h(r)$ first and second derivatives can be calculated, and inserting them into either (\ref{sol11}) or (\ref{sol12}), one obtains in both cases
\begin{eqnarray}
\rho &=& \frac{2(1-h)^4}{\kappa c_{\mu} h^2(1+h)^2} \left[\frac{ c_{\alpha}h(- 2\ln h + 4h - h^2 + c_{\beta})}{c_{\gamma}^2 (1+h)} \right.  \nonumber \\
&+& \left. 2 \frac{c^2}{c_f^2}(1-h)^2 \right]. \label{sol15}
\end{eqnarray}
Moreover, $h''$ given by (\ref{sol13}) can be inserted into (\ref{sol10}), therefore giving
\begin{eqnarray}
&&\frac{D'}{D} = \frac{h'}{2(1-h)} - \frac{h'}{2(1+h)} \nonumber \\
&+& \frac{c^2 c_{\gamma}^2}{2 c_{\alpha} c_f^2} \frac{(- 2\ln h + 4h - h^2)'}{(- 2\ln h + 4h - h^2 + c_{\beta})},  \label{sol16}
\end{eqnarray} 
which can be integrated as
\begin{equation}
D = c_d \frac{(- 2\ln h + 4h - h^2 + c_{\beta})^{c_{\delta}}}{\sqrt{(1-h)(1+h)}}, \label{sol17}
\end{equation}
where the integration constant combination $c^2 c_{\gamma}^2/2 c_{\alpha}c_f^2$ has been renamed $c_{\delta}$, and $c_d$ is a new integration constant.

Now (\ref{sol17}) inserted into (\ref{sol8}) allows one to integrate this equation as
\begin{equation}
k = \frac{c_f}{(1-h)^2}\left[ c_0 + c_k (- 2\ln h + 4h - h^2 + c_{\beta})^{c_{\delta} + \frac{1}{2}}\right], \label{sol18}
\end{equation}
where $c_k$ is a new integration constant into which a $\pm$ sign has been absorbed and where another integration constant has been added to $c_0$ without this constant notation being modified.
Finally, the last metric function $l$ emerges from inserting $f$, $k$, and $D$ into (\ref{D2}), which gives
\begin{eqnarray}
&&l = \frac{c_d^2}{c_f}\frac{(1-h)}{(1+h)}(- 2\ln h + 4h - h^2 + c_{\beta})^{2c_{\delta}} \nonumber \\
&-& \frac{c_f}{(1-h)^2} \left[c_0 + c_k (- 2\ln h + 4h - h^2 + c_{\beta})^{c_{\delta} + \frac{1}{2}}\right]^2. \label{sol19}
\end{eqnarray}

Now, inserting the above fully integrated metric function expressions into the field equations leads to three constraints upon the integration constants. The field equation (\ref{G22}) gives
\begin{equation}
c_{\delta} = \frac{1}{2}. \label{const1}
\end{equation}
Then, (\ref{G00}) implies
\begin{equation}
2  \frac{c_{\alpha} c_f^2 c_{\delta}}{c^2 c_{\gamma}^2}= 1. \label{const2}
\end{equation}
And finally, (\ref{G11}), or equivalently (\ref{G03}), yields
\begin{equation}
\left(c_{\delta} + \frac{1}{2}\right)^2 c_f^2c_k^2 = 2c_{\delta}c_d^2. \label{const3}
\end{equation}
The last field equation (\ref{G33}) confirms these constraints but does not impose new ones. Now, with (\ref{const1}) inserted, the two last constraint equations (\ref{const2}) and (\ref{const3}) can be written as
\begin{equation}
c_{\alpha} = \frac{c^2 c_{\gamma}^2}{c_f^2}, \label{const4}
\end{equation}
and
\begin{equation}
c_d^2 = c_f^2 c_k^2. \label{const5}
\end{equation}

Inserting (\ref{const1}), (\ref{const4}), and (\ref{const5}) into (\ref{sol17})--(\ref{sol19}), one obtains
\begin{equation}
D^2 = c_f ^2 c_k^2 \frac{(- 2\ln h + 4h - h^2 + c_{\beta})}{(1-h)(1+h)}, \label{sol20}
\end{equation}
\begin{equation}
k = \frac{c_f}{(1-h)^2}\left[ c_0 + c_k (- 2\ln h + 4h - h^2 + c_{\beta})\right], \label{sol21}
\end{equation}
\begin{eqnarray}
&&l = c_f c_k^2 \frac{(1-h)}{(1+h)}(- 2\ln h + 4h - h^2 + c_{\beta}) \nonumber \\
&-& \frac{c_f}{(1-h)^2} \left[c_0 + c_k (- 2\ln h + 4h - h^2 + c_{\beta})\right]^2. \label{sol22}
\end{eqnarray}

\subsection{Regularity conditions} \label{regcond}

The regularity conditions on the symmetry axis for metric (\ref{metric}) have already been displayed  in \cite{D06,CS20}. However, since they will be needed in the following, they are recalled briefly here. 

To ensure elementary flatness in the rotation axis vicinity, the norm $X$ of the Killing vector $\partial_{\phi}$ must satisfy \cite{S09}
\begin{equation}
\lim_{r \to 0} \frac{g^{\alpha \beta}X_{, \alpha}X_{,\beta}}{4X} = 1, \label{regcond1}
\end{equation}
where $X = g_{\phi \phi}$. Equations (\ref{metric}) and  (\ref{regcond1}) yield
\begin{equation}
\lim_{r \to 0} \frac{e^{-\mu} l'^2}{4l} = 1. \label{regcond2}
\end{equation}
The requirement that $g_{\phi \phi}$ vanishes on the axis implies
\begin{equation}
l \stackrel{0}{=} 0, \label{regcond3}
\end{equation}
where $ \stackrel{0}{=}$ means that the values are taken at $r=0$.

Since there cannot be singularities along the axis, it is imposed that, at this limit, spacetime tends to flatness, hence, the coordinates are scaled such that, for $r \rightarrow 0$, the  metric becomes
\begin{equation}
\textrm{d}s^2 = -\textrm{d}t^2 + 2\omega r^2 \textrm{d}t \textrm{d}\phi +\textrm{d}r^2 + \textrm{d}z^2 + r^2 \textrm{d}\phi^2, \label{axismetric}
\end{equation}
from which
\begin{equation}
f  \stackrel{0}{=} \textrm{e}^\mu  \stackrel{0}{=} 1, \qquad k \stackrel{0}{=} 0 \label{regcond4}
\end{equation}
follow, implying
\begin{equation}
D  \stackrel{0}{=} 0, \label{regcond5}
\end{equation}
and, from (\ref{regcond2}) and (\ref{axismetric}),
\begin{equation}
l'  \stackrel{0}{=} 0. \label{regcond6}
\end{equation}
Then, from the above and the requirement that the Einstein tensor components in  (\ref{G00})-(\ref{G33}) do not diverge, we have
\begin{equation}
f'  \stackrel{0}{=} k' \stackrel{0}{=} k'' - k'\frac{D'}{D} \stackrel{0}{=} 0. \label{regcond7}
\end{equation}

Inserting (\ref{sol2}), (\ref{sol4}) and (\ref{sol21}) into (\ref{regcond4}) one obtains
\begin{equation}
c_f  \stackrel{0}{=} (1-h)^2, \label{regcond8}
\end{equation}
\begin{equation}
c_{\mu}  \stackrel{0}{=} \frac{c_f}{h^2}, \label{regcond9}
\end{equation}
\begin{equation}
c_0 = 0, \label{regcond10}
\end{equation}
\begin{equation}
c_{\beta}  \stackrel{0}{=} 2 \ln h - 4h + h^2. \label{regcond11}
\end{equation}
All the other regularity conditions are verified provided (\ref{regcond10}) and (\ref{regcond11}) are satisfied.

Notice that (\ref{regcond11}) implies $c_{\beta} \neq -3$, otherwise it would impose $h \stackrel{0}{=} 1$ and $c_f$ given by (\ref{regcond8}) would be forced to vanish, which would rule out the whole solution.

\subsection{Junction conditions} \label{junct}

These conditions have also been displayed in \cite{D06,CS20} for metric (\ref{metric}). For completeness, and also since they will be partially needed further on, they are recalled here briefly. 
 
Outside the fluid cylinder, a vacuum solution to the field equations is needed. Since the system is stationary, the Lewis metric \cite{L32} will be used to represent such an exterior spacetime, and the Weyl class \cite{dS95} is chosen here for a real junction condition purpose. Its metric can be written as
\begin{equation}
\textrm{d}s^2=-F \textrm{d}t^2 + 2 K \textrm{d}t \textrm{d}\phi +\textrm{e}^M (\textrm{d}R^2 +\textrm{d}z^2) + L \textrm{d}\phi^2, \label{Wmetric}
\end{equation}
where
\begin{equation}
F= a R^{1 - n} - a \delta^2 R^{1 + n}, \label{W1}
\end{equation}
\begin{equation}
K = - (1 - ab\delta)\delta R^{1 + n} - ab  R^{1 - n}, \label{W2}
\end{equation}
\begin{equation}
\textrm{e}^M =  R^{(n^2 - 1)/2},  \label{W3}
\end{equation}
\begin{equation}
L = \frac{(1 - ab\delta)^2}{a} R^{1 + n} - ab^2 R^{1 - n}, \label{W4}
\end{equation}
with
\begin{equation}
\delta = \frac{c}{an}, \label{W5}
\end{equation}
where $a, b, c$, and $n$ are real constants. See \cite{D06} for comments about the respective coordinate systems inside and outside the fluid cylinder and \cite{dS95} for more details about the Lewis metric Weyl class.

In accordance with Darmois' junction conditions \cite{D27}, metric (\ref{metric}) and metric (\ref{Wmetric}) coefficients and their derivatives must be continuous across the $\Sigma$ surface,
\begin{equation}
f  \stackrel{\Sigma}{= }a_1 F, \quad k  \stackrel{\Sigma}{=} a_2 K, \quad \textrm{e}^\mu  \stackrel{\Sigma}{=} a_3 \textrm{e}^M, \quad l  \stackrel{\Sigma}{=} a_4 L, \label{W6}
\end{equation}
\begin{equation}
\frac{f'}{f}  \stackrel{\Sigma}{=} \frac{1}{R} + n \frac{\delta^2 R^n + R^{-n}}{\delta^2 R^{1 + n} - R^{1 - n}}, \label{W7}
\end{equation}
\begin{equation}
\frac{k'}{k}  \stackrel{\Sigma}{=} \frac{1}{R} + n \frac{(1 - ab\delta)\delta R^n - abR^{-n}}{(1 - ab\delta)\delta R^{1 + n} + abR^{1 - n}}, \label{W8}
\end{equation}
\begin{equation}
\mu' \stackrel{\Sigma}{=} \frac{n^2 - 1}{2R},  \label{W9}
\end{equation}
\begin{equation}
\frac{l'}{l}  \stackrel{\Sigma}{=} \frac{1}{R} + n \frac{(1 - ab\delta)^2 R^n + a^2b^2 R^{-n}}{(1 - ab\delta)^2 R^{1 + n} - a^2b^2R^{1 - n}}. \label{W10}
\end{equation}
The first fundamental form continuity imposes  (\ref{W6}) where the $a_1$, $a_2$, $a_3$ and $a_4$ constants can be transformed away by rescaling the coordinates, while (\ref{W7})--(\ref{W10}) are produced by the second fundamental form continuity. Hence, the above equations inserted into \cite{D06}'s Eq.(8) imply $P_r \stackrel{\Sigma}{=} 0$,  which is in perfect agreement with the present fluid equation of state which imposes $P_r=0$ everywhere. Moreover, the integration constant $c$ appearing in (\ref{sol6}) can be identified with the constant $c$ displayed in (\ref{W5}), which arises from the source stationarity producing the source vorticity \cite{MC98,dS02,D06}. This justifies the choice of the factor 2 for the definition of this integration constant in (\ref{sol6}).

Comments about other spacetime properties issuing from the above relations are displayed in \cite{D06,CS20}. They are not recalled here since they will not be needed for the present purposes.

\subsection{Final form of the solution} \label{final}

Constraints (\ref{regcond8})--(\ref{regcond11}) on the integration constants issuing from the regularity conditions and inserted into the metric functions as given in (\ref{sol4}), (\ref{sol2}), (\ref{sol21}), and (\ref{sol22}) lead to their final form, which can be written as
\begin{equation}
f =  \left(\frac{1-h_0}{1-h}\right)^2, \label{sol23}
\end{equation}
\begin{equation}
\textrm{e}^\mu = \left(\frac{1-h_0}{h_0}\right)^2 \left(\frac{h}{1-h}\right)^2, \label{sol24}
\end{equation}
\begin{equation}
k = (1-h_0)^2 c_k \frac{\left[ 2\ln \frac{h_0}{h} + 4 (h-h_0) - (h^2 - h_0^2)\right]}{(1-h)^2}, \label{sol25}
\end{equation}
\begin{eqnarray}
&&l =  (1-h_0)^2 c_k^2\left[ 2\ln \frac{h_0}{h} + 4 (h-h_0) - (h^2 - h_0^2)\right] \nonumber \\
&\times& \left\{\frac{1-h}{1+h} - \frac{\left[ 2\ln \frac{h_0}{h} + 4 (h-h_0) - (h^2 - h_0^2)\right]}{(1-h)^2}\right\}. \label{sol26}
\end{eqnarray}

Notice that since the constant $c_k$ appears as such in expression (\ref{sol25}) for $k$, and squared in expression (\ref{sol26}) for $l$, it can be absorbed into a rescaling of the $\phi$ coordinate according to the coordinate transformation $c_k \phi \rightarrow \phi$. Now, as it has been mentioned in (\ref{ranges}),  cylindrical symmetry imposes $0 \leq \phi \leq 2 \pi$. Therefore, such a rescaling is equivalent to setting $c_k = 1$ in (\ref{sol25}) and (\ref{sol26}), which will be adopted henceforth.

Expression (\ref{sol20}) for $D$ thus becomes
\begin{equation}
D^2 = (1-h_0)^4 \left[\frac{ 2\ln \frac{h_0}{h} + 4 (h-h_0) - (h^2 - h_0^2)}{(1-h)(1+h)}\right]. \label{sol27}
\end{equation}
The solution includes the equation for $h(r)$ that reads
\begin{eqnarray}
&&r = \frac{(1-h_0)^2}{c} \nonumber \\
&&\times \int_{h_0}^{h} \left\{\frac{1+u}{u^2(1-u)^3\left[ 2\ln \frac{h_0}{u} + 4 (u-h_0) - (u^2 - h_0^2)\right]}\right\}^\frac{1}{2} \textrm{d}u, \label{sol28}
\end{eqnarray}
where the integration constant $c_r$ has been absorbed into $h_0$ denoting the $h$ value on the symmetry axis, i.e., $h_0 \equiv h(r=0)$. Then, the expression for $\rho$ can be written as
\begin{eqnarray}
&&\rho = \frac{2c^2 h_0^2}{\kappa (1-h_0)^4} \frac{(1-h)^4}{h^2 (1+h)^2} \nonumber \\
&\times& \left\{\frac{h\left[ 2\ln \frac{h_0}{h} + 4 (h-h_0) - (h^2 - h_0^2)\right]}{(1+h)} +  2(1-h)^2 \right\}. \label{sol29}
\end{eqnarray}

\section{Physical properties of the solution}\label{physprop}

\subsection{Hydrodynamical scalars, vectors, and tensors} \label{tobetitled}

The timelike 4-vector $V_\alpha$ can be invariantly decomposed into three independent parts through the genuine tensor $V_{\alpha;\beta}$ as
\begin{equation}
V_{\alpha;\beta} = - \dot{V}_\alpha V_\beta + \omega_{\alpha \beta} + \sigma_{\alpha \beta}, \label{decomp}
\end{equation}
where 
\begin{equation}
 \dot{V}_\alpha = V_{\alpha;\beta} V^\beta, \label{accel}
\end{equation}
\begin{equation}
\omega_{\alpha \beta} = V_{\left[\alpha;\beta\right]} +  \dot{V}_{\left[\alpha\right.} V_{\left.\beta\right]}, \label{rotation}
\end{equation}
\begin{equation}
\sigma_{\alpha \beta} = V_{(\alpha;\beta)} +  \dot{V}_{(\alpha} V_{\beta)}. \label{shear}
\end{equation}
The three above quantities are called, respectively, the acceleration vector, the rotation or twist tensor, and the shear tensor. For the timelike 4-vector given by (\ref{r4velocity}), their nonzero components \cite{CS20} are
\begin{equation}
\dot{V}_1 = \frac{1}{2} \frac{f'}{f}, \label{dotV1a}
\end{equation}
which becomes, with (\ref{sol23}) inserted,
\begin{eqnarray}
\dot{V}_1 &=& \frac{c}{(1-h_0)^2} h \left\{\frac{(1-h)}{(1+h)}\left[2\ln \frac{h_0}{h} \right. \right. \nonumber \\
&+& \left. \left. 4 (h-h_0) - (h^2 - h_0^2)\right]\right\}^{\frac{1}{2}} \label{dotV1b}
\end{eqnarray}
and
\begin{equation}
2 \omega_{13} = - (2kv'+k'v). \label{omega13}
\end{equation}
From (\ref{dotV1a}), the acceleration vector modulus is therefore
\begin{equation}
\dot{V}^\alpha \dot{V}_\alpha = \frac{1}{4} \frac{f'^2}{f^2} \textrm{e}^{-\mu}, \label{modaccela}
\end{equation}
which becomes, with (\ref{sol23}) and (\ref{sol24}) inserted, and with (\ref{sol28}) differentiated with respect to $r$,
\begin{eqnarray}
\dot{V}^\alpha \dot{V}_\alpha &=& \frac{c^2 h_0^2}{(1-h_0)^6} \frac{(1-h)^3}{(1+h)} \nonumber \\
&\times& \left[ 2\ln \frac{h_0}{h} + 4 (h-h_0) - (h^2 - h_0^2)\right]. \label{modaccelb}
\end{eqnarray}
The rotation scalar $\omega$ defined by
\begin{equation}
\omega^2 = \frac{1}{2} \omega^{\alpha \beta}\omega_{\alpha \beta} \label{omega2def}
\end{equation}
follows as
\begin{equation}
\omega^2 = \frac{1}{4  \textrm{e}^{\mu}D^2}\left(k\frac{f'}{f}-k'\right)^2. \label{omega2a}
\end{equation}

Inserting  (\ref{sol6}) into (\ref{omega2a}), one obtains
\begin{equation}
\omega^2 = \frac{c^2}{f^2\textrm{e}^{\mu}}, \label{omega2c}
\end{equation}
which becomes, after inserting (\ref{sol23}) and (\ref{sol24}),
\begin{equation}
\omega^2 = \frac{c^2 h_0^2}{(1 - h_0)^6} \frac{(1 - h)^6}{h^2}. \label{omega2d}
\end{equation}

As already stressed in \cite{D06,CS20}, the shear tensor vanishes for any rigidly rotating fluid.

\subsection{Constant parameter interpretations} \label{parameters}

As usual in GR, the mathematical and therefore physical properties of the new solution displayed here through (\ref{sol23})--(\ref{sol29}) with $c_k=1$ depend strongly on the values exhibited by the two independent integration constants $h_0$ and $c$, which can be considered as solution parameters.

The $h_0$ interpretation is actually obvious from (\ref{sol28}). It is the $h(r)$ value on the $r=0$ symmetry axis. It has already been stressed in Sec. \ref{regcond} that the value $h_0 = 1$, i.e., $P_z \stackrel{0}{=} \rho$, is forbidden since, from regularity condition (\ref{regcond8}), it would imply $c_f = 0$ and thus rule out the entire solution.

Now, making $h = h_0$ into (\ref{omega2d}) gives $\omega^2 = c^2$. Hence, $|c|$ is the rotation scalar amplitude on the symmetry axis. Since this $c$ parameter is the same as the one appearing in the Lewis exterior spacetime metric, its absolute value can be interpreted in the vacuum framework as the amplitude of the interior gravitational source rotation scalar on the symmetry axis \cite{dS95}.

Moreover, it is obvious from (\ref{sol29}) that the larger $|c|$, the bigger $\rho$, for a given $\{h_0, r \rightarrow h(r)\}$ couple. And therefore, an increased (decreased) fluid vorticity on the axis induces, inside the $\Sigma$ boundary surface, an increased (decreased) energy density, and thus, an increased (decreased) pressure.

\subsection{Metric signature and sign constraints} \label{signcons}

To obtain the proper metric signature chosen here, i.e., $(-+++)$, every metric function as stated in (\ref{metric}) must be positive definite. The metric functions $f$ and $\textrm{e}^\mu$ as given by (\ref{sol23}) and (\ref{sol24}), respectively, are positive by construction.

Now, from (\ref{sol25}) with $c_k=1$, one can see that $k$ is positive, provided
\begin{equation}
p(h) \equiv 2\ln \frac{h_0}{h} + 4 (h-h_0) - (h^2 - h_0^2) > 0. \label{sign1}
\end{equation}
As regards $l$ given by (\ref{sol26}) with $c_k=1$, an analogous constraint implies, once (\ref{sign1}) is fulfilled,
\begin{equation} 
q(h) \equiv \frac{1-h}{1+h} - \frac{2\ln \frac{h_0}{h} + 4 (h-h_0) - (h^2 - h_0^2)}{(1-h)^2} > 0. \label{sign2}
\end{equation}
The expression (\ref{sol27}) for $D^2$ imposes, once (\ref{sign1}) is satisfied, 
\begin{equation}
\frac{1}{(1-h)(1+h)} >0. \label{sign3}
\end{equation}

The weak energy condition $\rho > 0$ implies from (\ref{sol29}) another constraint that reads
\begin{equation} 
\frac{h \left[2\ln \frac{h_0}{h} + 4 (h-h_0) - (h^2 - h_0^2)\right]}{1+h} + 2 (1-h)^2 > 0. \label{sign4}
\end{equation}

Finally, the logarithm function occurring in several places under the form $\ln h_0/h$ imposes
\begin{equation}
\frac{h_0}{h} >0. \label{sign5}
\end{equation}

It is straightforward to see that (\ref{sign3}) implies 
\begin{equation}
-1<h<1. \label{sign6}
\end{equation}

Now, from (\ref{sign5}), once the $h_0$ sign has been fixed, e. g., measured on the axis, $h$ must keep the same sign all along. Two cases can thus be distinguished:

(i) $h_0>0$, i.e., from  (\ref{sign6}), $0<h_0<1$, which implies, by continuity, $0<h<1$.

(ii)  $h_0<0$, i.e., from  (\ref{sign6}), $-1<h_0<0$, which implies, by continuity, $-1<h<0$.

\subsubsection{Case (i)}

A straightforward mathematical analyzis shows that the constraint $p(h)$ positive for any $h$ value such that $0<h<1$, as stated by (\ref{sign1}), imposes $0<h<h_2<1$, and, by continuity, $0<h_0<h_2<1$, with $h_2$ depending on $h_0$ according to
\begin{equation}
2 \ln \frac{h_0}{h_2} + 4(h_2 - h_0) - (h_2^2 - h_0^2) = 0. \label{sign7}
\end{equation}

Then, with $h$ and $p(h)$ positive, which is the case here, inequality (\ref{sign4}) is identically verified.

Another straightforward mathematical analysis shows that the constraint $q(h)$ positive for any $h$ value such that $0<h<h_2<1$, as stated by (\ref{sign2}), imposes $0<h_1<h<h_2<1$, and, by continuity, $0<h_1<h_0<h_2<1$, with $h_1$ depending on $h_0$ according to
\begin{equation}
(1-h_1)^3 - (1+h_1)\left[2 \ln \frac{h_0}{h_1} + 4(h_1 - h_0) - (h_1^2 - h_0^2) \right] = 0. \label{sign8}
\end{equation}

To summarize: Case (i), which is valid for $h_0>0$, implies $0<h_1<h_0<h_2<1$ and  $0<h_1<h<h_2<1$, with $h_1$ depending on $h_0$ through (\ref{sign8}), and $h_2$ depending on $h_0$ through (\ref{sign7}).

\subsubsection{Case (ii)}

An analogous method applies in this case. First, a straightforward mathematical analysis shows that $p(h)$ positive for any $h$ value such that $-1<h<0$ imposes $h_0<h$.

With $-1<h_0<h<0$, as it is therefore the case, inequality (\ref{sign4}) is identically satisfied.

Finally, a last straightforward mathematical analysis shows that $q(h)$ positive for any $h$ value such that $-1<h_0<h<0$ imposes $-1<h<h_3<0$, and, by continuity, $-1<h_0<h_3<0$, with $h_3$ depending on $h_0$ through
\begin{equation}
(1-h_3)^3 - (1+h_3)\left[2 \ln \frac{h_0}{h_3} + 4(h_3 - h_0) - (h_3^2 - h_0^2) \right] = 0. \label{sign9}
\end{equation}

To summarize: Case (ii), valid for $h_0<0$, implies $-1<h_0<h<h_3<0$,  with $h_3$ depending on $h_0$ through (\ref{sign9}).

Now, the $h<0$ range should be considered with caution as regards its physical interpretation, since it implies a negative pressure. However, such a feature, even though regarded as unphysical when standard fluids are considered, can emerge in some circumstances such as cosmological issues when, e.g., "dark energy" comes into play.

Notice moreover that in both cases (i) and (ii), and  in the units adopted here, i.e., $c=1$ and $8 \pi G = \kappa$, the pressure amplitude is smaller than that of the energy density.

\subsection{Singularities}

The solution displayed here exhibits three possible singularities.

A first one might occur for $h=+1$ where the whole metric function set diverges. However, this is not the case for the density $\rho$ which, from its expression (\ref{sol29}), is merely seen to vanish. The two Weyl scalar polynomial invariants $I$ and $J$ both vanish also at this limit. It is easy to convince oneself of this fact by inserting (\ref{sol23})--(\ref{sol27}), together with the present gauge choice $C_{0101}=C_{0202}=0 \Rightarrow C_{0303} = 0$, into the expressions for $I$ and $J$ displayed in \cite{CS20}'s (94)--(96). Even though a more complete analysis should be needed to conclude definitively, the question whether this locus might happen to be a mere coordinate singularity remains open.

For $h=-1$, $l$ is the only metric function which diverges, and therefore, so does $D$. However, at this location, $\rho$ also happens to diverge and would change sign if $h$ was allowed to reach values below $-1$, which is not the case as it has been shown in Sec. \ref{signcons}. Anyhow, this energy density behavior would imply a curvature singularity if such an $h$ value were reachable.

The third singularity might occur for $h=0$. Here, the density diverges and is no more defined if the $h$ sign happens to become different from the $h_0$ sign. We should be therefore confronted to another curvature singularity.

Now, notice that the three singularities occur for the three $h$ values which limit the definition intervals of $h$ distinguishing the positive and negative cases studied as (i) and (ii) in Sec. \ref{signcons}. They can therefore be excluded from the solution definition domain, as it has been proposed in Sec. \ref{signcons}, and thus, the solution becomes singularity-free.

In this case, the $P_z = 0$ spacetime is itself excluded from this solution class which therefore does not possess any dust limit.

\subsection{Behavior of the $h(r)$ function}

Differentiating (\ref{sol28}) with respect to the $r$ coordinate, one obtains
\begin{eqnarray}
&&h' = \frac{c}{(1-h_0)^2} \nonumber \\
&&\times \left\{\frac{h^2(1-h)^3\left[ 2\ln \frac{h_0}{h} + 4 (h-h_0) - (h^2 - h_0^2)\right]}{1+h}\right\}^\frac{1}{2}. \label{beh1}
\end{eqnarray}
This $h(r)$ function first derivative vanishes for $h=h_0$. For any other allowed value of $h$, as defined in Sec. \ref{signcons}, the (\ref{beh1}) right-hand side is nonzero and exhibits the same sign as that of the $c$ parameter. Once the $c$ sign is fixed, $h'$ keeps this sign all along from $r=0$ to $r= r_{\Sigma}$. 

Now, the two solution classes have to be examined separately since their properties differ.

\subsubsection{Case (i)}

In this case, where $h$ and $h_0$ are both positive, the $h(r)$ behavior depends on the $c$ sign. Actually,

(a) For $c<0$, which implies $h'$ negative, one have $h<h_0$ for every $r$ value, and therefore the function $h \equiv P_z/\rho$ is monotonously decreasing from $r= 0$ to $r= r_{\Sigma}$. This means that the $P_z$ amplitude decreases with respect to that of $\rho$ while going outward.

(b) For $c>0$, which implies $h'$ positive, $h>h_0$ for every $r$ value, and $h(r)$ is monotonically increasing from $r= 0$ to $r= r_{\Sigma}$. This means that the $P_z$ amplitude increases with respect to that of $\rho$ while going outward.

\subsubsection{Case (ii)}

In this case, it has been shown that $-1<h_0<h<0$, which implies, since the sign of $h'$ is fixed once for all, that the $h(r)$ function must be monotonically increasing from $h_0$, i.e., from $r=0$, up to $r= r_{\Sigma}$. Therefore, $h'>0$ follows, and, as a consequence, $c>0$. Here, of course, the $P_z$ amplitude increases with respect to that of $\rho$ while going outward.

\hfill

It is therefore interesting to notice that, in the positive pressure case, the sign of the $c$ parameter, which can be either positive or negative, influences the behavior of the pressure/energy density ratio, while this is not the case when the pressure is negative, since, then, $c$ is forced to be positive. Hence, for "ordinary" fluids, not only does $c$ measure the amplitude of the vorticity scalar on the symmetry axis, but its sign prescribes an important feature for the fluid, which might remind us of some kind of centrifugal-centripetal behavior.

\section{Conclusions}\label{concl}

Following the stationary cylindrical anisotropic fluid sourced interior spacetime investigations initiated in \cite{D06,CS20}, the rigidly rotating fluid case with particular equation of state $P_r = P_\phi=0$ has been examined deeper here. A field equation exact solution has thus been exhibited under the form of $h(r)$ functions for the metric and the density, with $h$ defined as $h(r) = P_z(r)/\rho(r)$, and an integral expression for $h$ as an $r$ function has been displayed.    Of course, as usual, this solution is valid in a given coordinate system which, however, has been chosen such as to allow a direct physical interpretation. To allow potential further uses for astrophysical purposes, it has been forced to satisfy the regularity conditions on the symmetry axis and has been matched to the Weyl class of the Lewis vacuum solution on a cylindrical $\Sigma$ hypersurface.

A number of physical and mathematical properties pertaining to this solution have been examined here such as the hydrodynamical scalars, vectors, and tensors which have been obtained as functions of $h(r)$. Singularities have been identified and discussed. It has been shown that they can, however, be excluded from the solution definition intervals, yielding thus singularity-free spacetimes.

The two independent parameters exhibited by this solution have been examined. The first one, $h_0$, corresponds to the value of the $h$ function on the symmetry axis. Its sign defines two different solution classes which have been displayed and discussed in Sec. \ref{signcons}. The second parameter, $c$, has been shown to represent the fluid vorticity amplitude on the axis. It corresponds to the $c$ parameter of the exterior Lewis-Weyl vacuum which thus inherits a precise confirmation of its previous vorticity interpretation \cite{dS95}. Moreover, in the case when the fluid pressure is positive, its sign determines the $h(r)$ increasing or decreasing property, while, in the negative pressure case, the $c$ sign is forced to be positive, while $h(r)$ is increasing from the axis up to the boundary surface.

As to allow a glimpse at such a simple fluid spacetime possible physical interpretations, the following results found in the literature can be considered. 

In 1992, Apostolatos and Thorne \cite{A92} have shown, using the simple analytic example of a thin cylindrical shell made of counterrotating dust particles, that an infinitesimal amount of rotation can halt the relativistic gravitational collapse of a pressure-free cylindrical body. Even though the here-displayed solution exhibits one nonzero pressure component, it could be put forward that since this component is axially directed, its influence on radial collapse might be negligible. Hence, the conjecture that such spacetimes might represent the final stage of some collapsing fluids might be contemplated, even though with some caution.

In 1996, Opher {\it et al.} \cite{O96} showed that, in rigidly rotating stationary cylindrical dust, test particle confinement occurs in the radial direction, while motion in the axial direction is free. They thus proposed that such a behavior might be relevant to extragalactic jet formation. Their arguments went as follows. The gravitational field produced by jets is usually negligible compared with the one produced by the matter at the galaxy centers. Thus, to first-order approximation, it is sufficient to model those jets as made of test particles. Also, since almost all the galaxies are rotating, as a first step, one can model a galaxy center by a rotating cylinder. This approximation seems reasonable to these authors as long as the gravitational field in the middle of the rotating galaxy is concerned, though they admit that it is indeed highly simplified. Assuming that such a model can capture some essence of physics, they claim that the confinement can be related to the jets. 

Bringing together such a confinement property and Apostolatos and Thorne's result might suggest some practical uses of a solution of the kind presented here. Actually, fully realistic models are very seldom encountered in GR, while exact solutions have often been used as very fruitful approximate ones.

Therefore, in any case, this simple class of anisotropic fluid source for cylindrically symmetric spacetimes, since it is a first attempt to deal with anisotropic pressure in such a framework, must be considered as progress for the Einstein field equation understanding from both a mathematical and a physical point of view. Actually, as stressed in \cite{G09}, "much can be learned about the character of gravitation and its effects by investigating particular idealised examples."

Moreover, the solution displayed here might also be viewed as a first step toward a future generalization to less simple anisotropic cases, possibly obtained by relaxing one or either simplifying assumptions made here, provided such analytic solutions happen to exist, of course.

\section*{Acknowledgments}

The author wants to thank the anonymous referee for accurate remarks and suggestions which has allowed her to improve this paper over a previous version.


\begin{thebibliography}{10}

\bibitem {LC19} T. Levi-Civita, $ds^2$ einsteiniani in campi newtoniani. IX: L'analogo del potenziale logaritmico, Rend. Accad. Lincei {\bf 28}, 101 (1919).
\bibitem {L24} C. Lanczos, \"Uber eine station\"are Kosmologie in Sinne der Einsteinischen Gravitationstheorie, Z. Phys. {\bf 21}, 73 (1924) [Gen. Relativ. Gravit. {\bf 29}, 363 (1997)].
\bibitem {L32} T. Lewis, Some special solutions of the equations of axially symmetric gravitational fields, Proc. R. Soc. A {\bf 136}, 176 (1932).
\bibitem {vS37} W. J. van Stockum, The gravitationl field of a distribution of particles rotating about an axis of symmetry, Proc. R. Soc. Edinburg A {\bf 57}, 135 (1937).
\bibitem {C19} M.-N. C\'el\'erier, R. Chan, M. F. A. da Silva, and N. O. Santos, Translation in cylindrically symmetric vacuum, Gen. Relativ. Gravit. {\bf 51}, 149 (2019).
\bibitem {G09} J. B. Griffiths and J. Podolsk\'y, {\it Exact spacetimes in Einstein's General Relativity,} Cambridge Monographs on Mathematical Physics (Cambridge University Press, Cambridge, England, 2009)
\bibitem {S09} H. Stephani, D. Kramer, M. MacCallum, C. Honselaers, and E. Herlt, {\it Exact Solutions to Einstein's Field Equations,} Cambridge Monographs on Mathematical Physics (Cambridge University Press, Cambridge, England, 2009)
\bibitem {B20} K. A. Bronnikov, N. O. Santos, and A. Wang, Cylindrical systems in general relativity, Classical Quantum Gravity {\bf 37}, 113002 (2020).
\bibitem {D06} F. Debbasch, L. Herrera, P. R. C. T. Pereira, and N. O. Santos, Stationary cylindrical anisotropic fluid, Gen. Relativ. Gravit. {\bf 38}, 1825 (2006).
\bibitem {CS20} M.-N. C\'el\'erier and N. O. Santos, Stationary cylindrical anisotropic fluid and new purely magnetic GR solutions,  Phys. Rev. D {\bf 102}, 044026 (2020).
\bibitem {dS95} M. F. A. da Silva, L. Herrera, F. M. Paiva, and N.O. Santos, The parameters of the Lewis metric for the Weyl class, Gen. Relativ. Gravit. {\bf 27}, 859 (1995).
\bibitem {D27} G. Darmois, Les \'equations de la gravitation einsteinienne, {\it M\'emorial des Sciences Mathématiques} fasc. XXV (Gauthier-Villars, Paris, 1927).
\bibitem {MC98} M. A. H. MacCallum and N. O. Santos, Stationary and static cylindrically symmetric Einstein spaces of the Lewis form, Classical Quantum Gravity {\bf 15}, 1627 (1998).
\bibitem {dS02} M. F. A. da Silva, L. Herrera, N. O. Santos, and A. Z. Wang, Rotating cylindrical shell source for Lewis spacetime, Classical Quantum Gravity {\bf 19}, 3809 (2002).
\bibitem {A92} T. A. Apostolatos and K. S. Thorne, Rotation halts cylindrical, relativistic gravitational collapse, Phys. Rev. D {\bf 46}, 2435 (1992).
\bibitem {O96} R. Opher, N. O. Santos, and A. Wang, Geodesic motion and confinment in van Stockum space-time, J. Math. Phys. (N.Y.) {\bf 37}, 1982 (1996).

\end{thebibliography}
\end{document}